\documentstyle[12pt,graphicx,caption2]{article}

\begin{document}

\title{Qualitative aspects of the models of $n\bar{n}$ transition in medium}
\author{V.I.Nazaruk\\
Institute for Nuclear Research of RAS, 60th October\\
Anniversary Prospect 7a, 117312 Moscow, Russia.}

\date{}
\maketitle
\bigskip

\begin{abstract}
We briefly outline the present state of the $n\bar{n}$ transition problem
and compare two alternative models.
\end{abstract}

\vspace{5mm}

\vspace{5mm}

\vspace{1cm}

*E-mail: nazaruk@inr.ru

\newpage
\setcounter{equation}{0}
\section{Introduction}
At present two models of $n\bar{n}$ transitions in medium shown in Figs. 1 and 2 (model
1 and 2, respectively) are treated. The models give radically different results. This is
because the process under study is extremely sensitive to the model and so we focus on the 
physics of the problem.

In the standard calculations of $ab$ oscillations in the medium [1-3] the interaction of 
particles $a$ and $b$ with the matter is described by the potentials $U_{a,b}$ (potential 
model). ${\rm Im}U_b$ is responsible for loss of $b$-particle intensity. In particular, 
this model is used for the $n\bar{n}$ transitions in a medium [4-10] followed by 
annihilation:
\begin{equation}
n\rightarrow \bar{n}\rightarrow M,
\end{equation}
here $M$ are the annihilation mesons.

In [9] it was shown that one-particle model mentioned above does not describe the
total $ab$ (neutron-antineutron) transition probability as well as the channel corresponding 
to absorption of the $b$-particle (antineutron). The effect of final state absorption
(annihilation) acts in the opposite (wrong) direction, which tends to the additional
suppression of the $n\bar{n}$ transition. The $S$-matrix should be unitary.

In [11] we have proposed the model based on the diagram technique (see Fig. 1). This model 
(later on referred to as the model 1) does not contain the non-hermitian operators. For 
deuteron this calculation was repeated in [12]. However, in [13] it was shown that this model
is unsuitable for the problem under study: the neutron line entering into the $n\bar{n}$ 
transition vertex should be the {\em wave function} (see Fig. 2), but not the propagator, 
as in the model 1. For the problem under study this fact is crucial. It leads to 
the cardinal error and so we abandoned the model 1. The model shown in Fig. 2a (model 2)
has been proposed [10,13]. 

\section{Model 1}
Consider now the model 1 [11]. The Hamiltonian of $n\bar{n}$ transition is [6]
\begin{equation}
{\cal H}_{n\bar{n}}=\epsilon \bar{\Psi }_{\bar{n}}\Psi _n+H.c.
\end{equation}
Here $\epsilon $ is a small parameter with $\epsilon =1/\tau _{n\bar{n}}$, where $\tau 
_{n\bar{n}}$ is the free-space $n\bar{n}$ oscillation time.

\begin{figure}[h]
  {\includegraphics[height=.25\textheight]{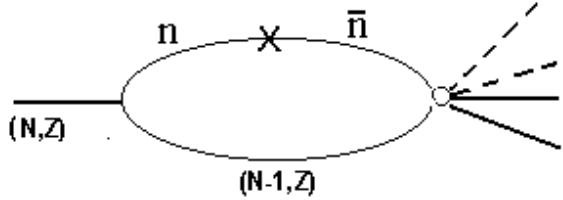}}
  \caption{Model 1 for the $n\bar{n}$ transition in the nuclei followed by annihilation.}
\end{figure}

Denote: $A=(N,Z)$, $B=(N-1,Z)$ are the initial and intermediate nuclei, $M$ is the
amplitude of virtual decay $A\rightarrow n+(A-1)$, $M_a^{(n)}$ is the amplitude of 
$\bar{n}B$ annihilation in $(n)$ mesons, $E_n$ is the pole neutron binding energy; $m$, 
$m_A$, $m_B$ are the masses of the nucleon and nuclei $A$ and $B$, respectively. The 
amplitude is  
\begin{equation}
M^{(n)}=-\frac{im^2m_B}{2\pi ^4}\epsilon 
\int d{\bf q}dE \frac{M(q)M_a^{(n)}(m_A)}{({\bf q}^2-2mE-i0)^2[{\bf q}^2+2m_B
(E+E_n)-i0)]}.
\end{equation}
For the process probability $W_1(t)$ one obtains
\begin{equation}
W_1(t)=\frac{\epsilon ^2}{6E_n^2}\Gamma t,
\end{equation}
where $\Gamma $ is the width of $\bar{n}B$ annihilation. The lower limit on the free-space 
$n\bar{n}$ oscillation time obtained by means of the distributions $W_1(t)$ is
\begin{equation}
\tau ^1_{{\rm min}}=(0.7-2.5)\cdot 10^{8}\; {\rm s}
\end{equation}

The main drawbacks of the model are as follows:

1) The model does not reproduce the $n\bar{n}$ transitions in the medium and vacuum. If the 
neutron binding energy goes to zero, Eq. (4) diverges (see also Eqs. (15) and (17) of 
Ref. [12]). 

2) Since the model is formulated in the momentum representation, it does not describe the 
coordinare-dependence, in particular the loss of particle intensity due to absorption.
This means that the model is rather crude and has a restricted range of applicability. 

3) The model does not contain the infrared singularity for any process including the 
$n\bar{n}$ transition, whereas it exists for the processes in the medium and vacuum (see
[10,13] and Sect. 3 of this paper). This brings up the question: Why? The answer is that 
for the propagator the infrared singularity cannot be in principle since the particle is 
virtual: $p_0^2\neq m^2+{\bf p}^2$. Due to this the model is infrared-free.

On the other hand, the neutron propagator arises owing to the vertex of virtual decay 
$A\rightarrow n+(A-1)$. However, in the interaction Hamiltonian {\em there is no term} which
induces the virtual decay $A\rightarrow n+(A-1)$ and so the model 1 cannot be obtained
from fundamental Hamiltonians. The diagram technique has been developed and adapted to the 
direct type reactions. In particular, it was applied by us for the calculation of knock-out 
reactions and $\bar{p}$-nuclear annihilation [14]. The approach is very handy, useful and 
simple since it is formulated in the momentum representation. The price of this simplicity 
is that its applicability range is {\em restricted}. On the other hand, the process under 
study is extremely sensitive to the value of antineutron self energy and the description 
of neutron state. The problem is unstable [15].

\section{Model 2}
In this section the more rigorous approach is considered. Model 2 shown in Fig. 2 
corresponds to the standard formulation of the problem: the $|in>$-state is described by 
the eigenfunctions of unperturbed Hamiltonian. In the case of diagram 2a, this is the 
neutron plane wave:
\begin{equation}
n_p(x)=\Omega ^{-1/2}\exp (-ipx),   
\end{equation}
$p=(\epsilon _n,{\bf p}_n)$, $\epsilon _n={\bf p}_n^2/2m+U_n$, where $U_n$ is the neutron 
potential. In the case of diagram 2b, this is the wave function of bound state. For the
nucleus in the initial state we take the one-particle shell model.

\begin{figure}[h]
  {\includegraphics[height=.25\textheight]{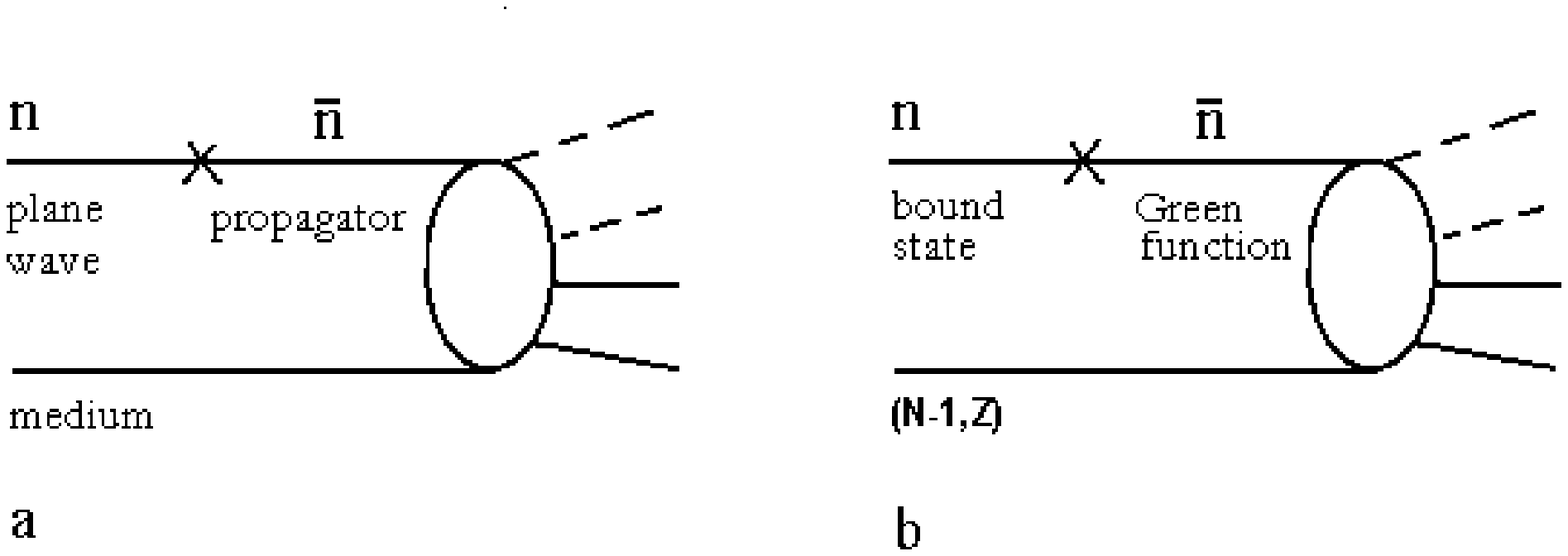}}
  \caption{Model 2 for the $n\bar{n}$ transition in the medium ({\bf a}) and nuclei ({\bf b})
followed by annihilation.}
\end{figure}

The interaction Hamiltonian is given by
\begin{equation}
{\cal H}_I={\cal H}_{n\bar{n}}+{\cal H},
\end{equation}
where ${\cal H}$ is the hermitian Hamiltonian of $\bar{n}$-medium interaction in the case
of Fig. 2a and Hamiltonian of $\bar{n}$-nuclear interaction in the case of Fig. 2b.

Consider now the diagram 2a. We show the result sensitivity to the details of the model.
If we use the general definition of amplitude of antineutron annihilation in the 
medium $M_a^m$ which is given by
\begin{eqnarray}
<\!f\!\mid T^{\bar{n}}\mid\!0\bar{n}_p\!>=N(2\pi )^4\delta ^4(p_f-p_i)M_a^m,\nonumber\\
T^{\bar{n}}=T\exp (-i\int_{-\infty}^{\infty}dtH(t))-1
\end{eqnarray}
($\mid\! 0\bar{n}_p\!>$ is the state of the medium containing the $\bar{n}$ with the 
4-momentum $p$, $N$ includes the normalization factors) then the amplitude corresponding to 
Fig. 2a diverges:
\begin{eqnarray}
M_a=\epsilon G_0^mM_a^m,\nonumber\\
G_0^m=\frac{1}{\epsilon _n-{\bf p}_n^2/2m-U_n}\sim \frac{1}{0}.
\end{eqnarray}
This is infrared singularity conditioned by zero momentum transfer in the $n\bar{n}$ 
transition vertex. We use the approach with a finite time interval which is
infrared-free. For the probability of the process (1) we get [10,13]
\begin{equation}
W_a(t)\approx W_f(t)=\epsilon ^2t^2,
\end{equation}
where $W_f$ is the free-space $n\bar{n}$ transition probability.

In the case considered above the amplitude $M_a^m$ involves all the $\bar{n}$-medium 
interactions followed by annihilation including the antineutron rescattering in the initial 
state. In principle, the part of this interaction can be included in the antineutron
propagator. Then the antineutron self-energy $\Sigma $ is generated and
\begin{equation}
G_0^m\rightarrow  G^m=\frac{1}{\epsilon _n-{\bf p}_n^2/2m-U_n-\Sigma }=-\frac{1}{\Sigma }.
\end{equation}
The process amplitude
\begin{equation}
M_b=\epsilon G^mM'_a
\end{equation}
is non-singular. The process probability is found to be [15]
\begin{equation}
W_b(t)\approx \frac{\epsilon ^2}{\Sigma ^2}\Gamma t.
\end{equation}
The result differs from (10) fundamentally. Using the distributions $W_b(t)$ and $W_a(t)$, 
for the lower limits on the free-space $n\bar{n}$ oscillation we get
\begin{equation}
\tau ^b_{{\rm min}}=(3.5-7.5)\cdot 10^{8}\; {\rm s}
\end{equation}
and 
\begin{equation}
\tau ^a_{{\rm min}}=10^{16}\; {\rm yr},
\end{equation}
respectively. So the realistic limit $\tau _{{\rm min}}$ can be in the range
\begin{equation}
\tau ^a_{{\rm min}}>\tau _{{\rm min}}>\tau ^b_{{\rm min}}.
\end{equation}
This is because the amplitude $M_a$ is in the peculiar point. The result is extremely 
sensitive to the value of antineutron self-energy $\Sigma $ as well as the description of 
initial neutron state and the value of momentum transfered in the $n\bar{n}$ transitions 
vertex. This is a problem of great nicety. Further investigations are desirable [15]. 
(Note that $\tau ^b_{{\rm min}}$ exceeds $\tau ^1_{{\rm min}}$ by a factor of five.)

\section{Discussion}
Since the operator (2) acts on the neutron state, in the model 1 the vertex of virtual 
decay $A\rightarrow n+(A-1)$ is introduced because one should separate out the neutron
state. This scheme is artificial since in the interaction Hamiltonian {\em there is no term}
which induces the virtual decay $A\rightarrow n+(A-1)$. Alternative method is given by the
model 2 which does not contain the above-mentioned vertex. The shell model used in the
$|in>$-state of the model 2 has no need of a commentary. Taking into account the result
sensitivity to the description of neutron state, we argue that the model 1 is inapplicable
to the problem under study.

In the recent manuscript [16] the previous calculations [11,12] have been repeated. We 
should comment the main statement of Sect. 5 of [16] (for more details, see [17]) since it 
is connected with the principal distinction between models 1 and 2 given above. The 
author writes: "If the infrared divergence takes place for the process of $n\bar{n}$ 
transitions in nucleus, it should take place also for the nucleus form-factor at zero 
momentum transfer".

It cannot take place in principle since the model 1 is infrared-free for any process 
including the $n\bar{n}$ transition. As shown above, this model is inapplicable for the 
$n\bar{n}$ transitions in nuclei.

In the footnote on the page 11 it is argued that "If the neutron line entering 
into the $n\bar{n}$ transition vertex is the wave function then this means that new rules 
are proposed. These "new rules" should allow to reproduce all the well known results of 
nuclear theory". In other words, in odder to reproduce all the well known results of
nuclear theory, the line entering into vertex should be the propagator.

This statement is at least strange. For example, in the distorted-wave impulse
approximation the nucleon interacting with incident particle is described by wave function
but not the propagator. The neutron line entering into the vertex is the propagator in the 
intermediate state only. 

It should be emphasised that oscillations in the vacuum and gas, in particular 
oscillations of ultracold neutrons in storage vessels [18] are considered as well. These 
processes are described in the framework of the model shown in Fig. 2a [15]. As noted above, 
they cannot be reproduced by means of model 1 in principle.

In the next paper the calculation of the process shown in Fig. 2b will be presented.

\end{document}